\documentclass[aps,pre,twocolumn,showpacs,floats]{revtex4}

\usepackage{graphicx}

\begin{document}

\title{Growing networks with local rules:\\ preferential attachment, clustering hierarchy and degree
correlations}

\author{Alexei V\'azquez}

\affiliation{Department of Physics, University of Notre Dame, Notre Dame, IN 46556, USA}

\date{\today}

\begin{abstract}

The linear preferential attachment hypothesis has been shown to be quite successful to explain the
existence of networks with power-law degree distributions. It is then quite important to determine if this
mechanism is the consequence of a general principle based on local rules. In this
work it is claimed that an effective linear preferential attachment is the natural outcome of growing
network models based on local rules. It is also shown that the local models offer an explanation to other
properties like the clustering hierarchy and degree correlations recently observed in complex networks.
These conclusions are based on both analytical and numerical results of different local rules, including
some models already proposed in the literature.

\end{abstract}


\maketitle

\section{Introduction}

In the last few years there has been a great interest in the study of networks, with particular emphasize
on the following properties: small world effect \cite{ws98,w99}, power law degree
distribution \cite{ab01a,dm002b} and more recently degree correlations \cite{pvv01,vpv02a,n02a} and
clustering hierarchy \cite{vpv02a,vpv02b,rb02}. This explosion has been possible thanks to the increase of
available network maps offering the graph
representation for a wide variety of systems with sizes ranging from hundred to billions of nodes. Examples
include technological networks such as the physical Internet
\cite{nlanr,lucent,scan,caida,fff99,gt00c,cmp00,pvv01,vpv02a,yjb01}, the World Wide Web
\cite{ajb99,bajb00a,bkm00}, electronic mail \cite{emb02,nfb02}, and electronic circuits \cite{cjs01},
biological networks such as the protein-protein interaction network \cite{ugc00b,itm00,ico01,jmbo01,w01},
metabolic paths \cite{jtaob00,wf}, and food webs \cite{cgn01,sm01}, social networks represented by the
citation graph \cite{ls98,r98,v01a}, scientific collaboration webs \cite{n00f,n00,jgn01,bjnr01a}, sexual
relations \cite{leasa01}, among others.

In particular metrics like the degree (the number of edges incident to a vertex), the minimum path
distance between pairs of vertices and the clustering coefficient (the fraction of edges among the neighbors
of a vertex) have attracted the attention of the physics community. Watts and Strogatz \cite{ws98,w99} have
shown that, in general, real networks are characterized by a small average minimum path distance and a
large clustering coefficient that together are named as the {\em small world effect}. The name comes
from the fact that we can reach every vertex in the graph crossing a small number of edges. Moreover,
Barab\'asi and collaborators \cite{ba99,baj99} have pointed out that many real networks are also
characterized by power law degree distributions, giving an appreciable probability to observe high degree
vertices. A more exhaustive analysis reveals that, in addition to power laws, truncated power laws and
exponential distributions are also observed \cite{asbs00}.

Barab\'asi and Albert (BA) proposed 
a mechanism that explains the origin of power law degree
distributions \cite{ba99}. This mechanism is based on two fundamental properties of a wide class of
real networks, their growing nature and the existence of a preferential attachment: new vertices added to
the graph are attached preferentially to high degree vertices. In particular a linear preferential
attachment, where the probability to get connected to a vertex is proportional to its degree, leads to
power law degree distributions. The preferential attachment mechanism can be generalized in different
ways. A sub-linear preferential attachment leads to bounded degree distributions while a super-linear one
yields graphs with a single hub connected to almost any other vertex \cite{krl00,kr01}. The power laws
can be also truncated after the introduction of other ingredients such as aging \cite{dm00}, bounded
capacity
\cite{asbs00} or limited information \cite{mbsn02}. Moreover, the introduction of quenched \cite{bb00a}
and annealed \cite{dms02,vfmv01} disorder leads to logarithmic corrections and multi-fractal scaling,
respectively.

The BA model provides a general mechanism to obtain power law
degree distributions in growing networks. If one consider other measures like the clustering coefficient
then one may conclude that this model is still insufficient to describe real graphs. However, we should
not focus on the detailed properties of the model but on its philosophy. That is, if we assume that there
is a growing tendency of the network and an effective linear preferential attachment then we obtain a
scale-free degree distribution. Actually, this effective preferential attachment have been measured in
different real graphs, including the Internet \cite{jnb01a,pvv01} and a variety of scientific
collaboration webs \cite{jnb01a,n01b,bjnr01a}, supporting the hypothesis of a linear attachment rate. With
regard to the other topological properties, we can construct many models with different clustering
coefficients, minimum path distances, and other metrics \cite{bollobas02}. However, the origin of the
ubiquity of the linear preferential attachment is not clear yet.

The topology of real networks is also characterized by degree correlations \cite{pvv01,n02a} and
clustering hierarchy \cite{vpv02a,rb02}.  Moreover, these correlations influence the
behavior of models defined on top of these graphs, as it has been recently shown in Refs.
\cite{n02a,bp02,bl02,vw02,bpv02,vm02}.
Growing network models with global evolution rules, like the
BA model, exhibit degree correlations. For instance, non-trivial degree correlations
has been obtained in the linear preferential attachment model \cite{kr01} and in a growing network model
without any preferential attachment \cite{chk01}. However, the degree correlations obtained in those
global models are not sufficiently strong to account for the features observed in real graphs. New models
giving a better representation of real graphs are starting to emerge \cite{ke02,hk02,rb02}. In addition to
the numerical simulations some analytical treatments have shown that power-law degree distributions and
clustering hierarchy are obtained as an outcome of these models \cite{rb02,sak02,vbmpv02,guido1,guido2,kang}. 
However, a general principle based on local rules is still missing.

In this work different {\em local} mechanisms that lead to graphs with power-law degree
distributions, degree correlations and clustering hierarchy are studied. The term {\em local} means
that we will investigate evolution rules that involve a vertex and its neighbors. As it will be
shown the preferential attachment, the inverse proportionality between the average clustering coefficient
and the vertex degree, and degree correlations are common features of growing graph models built by local
rules. The general principles behind these features are also determined.

The paper is organized as follows. In the first section the motivation for this work is presented. It is shown that, in
addition to power-law degree distributions, clustering hierarchy and degree correlations are common features of real
networks. Then in the following sections three different models based on local rules are presented. In all cases both
analytical and numerical evidence is provided. In particular, in Section \ref{sec:rw} a walk model is proposed as a
mechanism for searchable networks such as the WWW and the citation network. Then in Section \ref{sec:cnn} a model for
social network evolution is analyzed, based on the existence of potential connections between the neighbors of a vertex.
Finally, in Section \ref{sec:copy} we study models with duplication or replication of its vertices. The common patterns
observed on these models are summarized in the concluding Section \ref{sec:conc}.

\section{Correlations and hierarchy in real graphs}
\label{sec:exp}

In this section we study correlations in some real graphs. In particular we consider five different
networks here denoted by Router, AS, WWW, Gnutella, PIN and Math. In all cases the graph is obtained by
representing the ``relevant'' units of the system by vertices and their interactions or relations by
edges. In some cases, multiple graph representations of the same system can be obtained.
Router: is the router level graph representation of the Internet, where each vertex represents
a router and each edge represents a physical connection among them. AS: is the {\em
autonomous system} (AS) representation of the Internet, where each vertex represents an AS or service
provider and each edge represents a peer relation among them. WWW: is the graph representation
of the WWW, where each vertex represents a web page and each directed edge a hyperlink from one page to
another. Here we will consider the directed edges as undirected. Gnutella: is the graph
representation of the peer-to-peer network of the same name, where each vertex represents a user and each
edge a peer relation among them. PIN: is the graph representation of the protein interaction
network, where each vertex represents a protein and each edge an interaction among them. Math: is the
graph representation of the mathematical co-autorship network, where each vertex represents an author an
each edge the existence of at least one common publication among them.

In general, real networks are correlated and correlations may have different origins. Let
us consider the example of the Internet. Due to installation costs, the Internet has been designed
with a hierarchical structure. This hierarchy can be schematically divided in international
connections, national backbones, regional networks, and local area networks. Vertices providing
access to international connections or national backbones are off course on top level of this
hierarchy, since they make possible the communication between regional and local area networks.
Moreover, in this way, a small average minimum path distance can be achieved with a small
average degree. This hierarchical structure will introduce some correlations in the network
topology. For instance, it is expected that vertices with high degrees are connected to vertices with
low degrees.

On the contrary, in social networks well connected people tend to be connected with well
connected people \cite{n02a}. Let us take the example of the scientific co-authorship graph. A
scientist writing a lot of papers have in general a larger probability to write a paper with
another scientist who has also a lot of papers, than with one with a few papers. In fact, if
$F_i$ is the number of papers of scientist $i$ and $F_i\ll N$ then the probability that two
scientist write a paper together is roughly $F_iF_j/N$. Now, $F_i$ is in general a monotonic
increasing function of the scientist degree $d_i$ (number of collaborators) and, therefore,
scientists with a high degree will have a better chance to make a new article together, {\em
\i.e.} to be connected.

\begin{figure*}
\begin{center}
\includegraphics[width=6in]{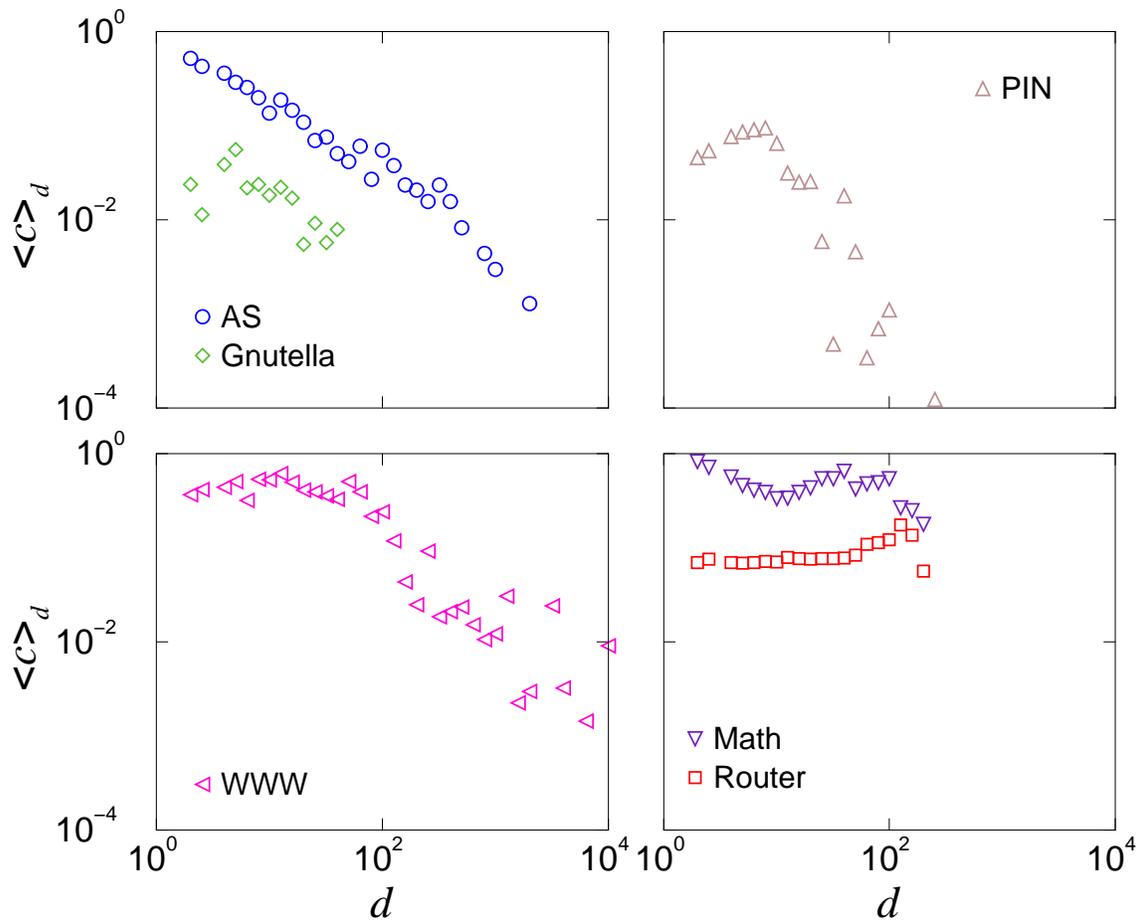}
\end{center}

\caption{Clustering coefficient as a
function of the vertex degree for some real graphs. AS and Router are the autonomous system \cite{nlanr} and router
\cite{scan} level graph representations of the Internet, respectively. WWW a sub-graph of the WWW network, a data set
collected by the Notre Dame group of Complex Networks (http://www.nd.edu/\symbol{126}networks). Gnutella is the Gnutella
peer to peer network, provided by Clip2 Distributed Search Solutions. PIN if the protein-protein interaction graph of {\em
Saccharomices Cerevisiae} as obtained from two hybrid experiments \cite{ico01}. Math is the co-authorship graph obtained
from all relevant journals in the field of mathematics and published in the period 1991-1998 \cite{bjnr01a}.}

\label{fig2:cls}
\end{figure*}

To investigate these correlations it has been proposed to analyze the clustering coefficient and the
nearest neighbor average connectivity as a function of the vertex degree \cite{pvv01,vpv02a}. The clustering
coefficient is the average probability that two neighbors $l$ and $m$ of a vertex $i$ are connected. In
terms of the adjacency matrix ($J_{ij}=1$ if vertices $i$ and $j$ are connected and 0 otherwise) the
clustering coefficient is defined as the conditional probability that if $J_{il}J_{im}=1$ then $J_{lm}=1$.
Thus, it measures in some way the existence of three-point correlations in the adjacency matrix. The
clustering coefficient $c_i$ is then defined as the ratio between the number of edges $e_i$ among the
$d_i$ neighbors of a given vertex $i$ and its maximum possible value, $d_i(d_i-1)/2$, {\em i.e.}
\begin{equation}
c_i = \frac{2e_i}{d_i(d_i-1)}.
\label{c_i}
\end{equation}
The average clustering coefficient $\left< c\right>$ is the average of $c_i$ over all vertices in the
graph. It provides a measure of how well the neighbors
of a vertex are locally interconnected. In Refs.~\cite{ws98,w99} it have been shown that the clustering
coefficient of many graphs
representing real systems is orders of magnitude larger than the one expected for a random graph and,
therefore, they are far from being random. Further information can be extracted if one compute it as a
function of the vertex degree \cite{vpv02a}.

\begin{figure*}
\begin{center}
\includegraphics[width=6in]{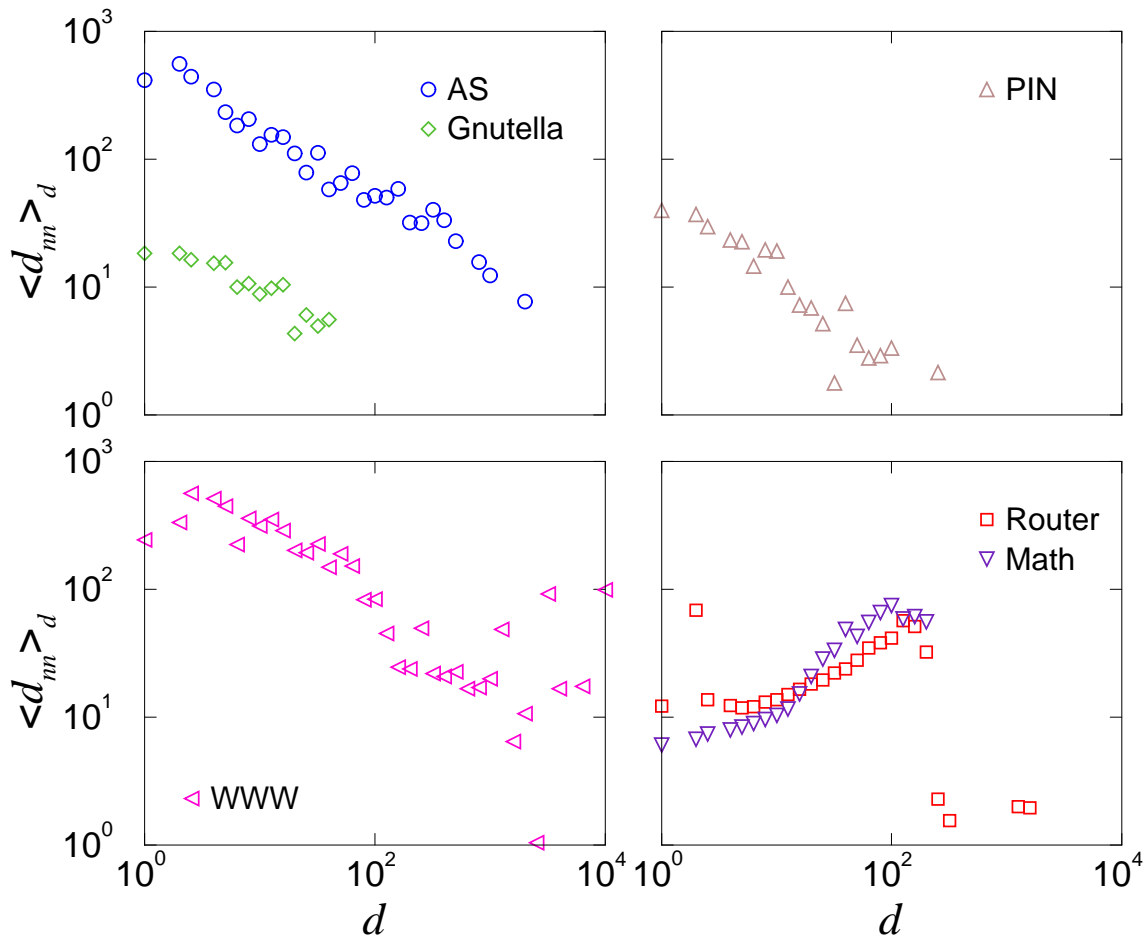}
\end{center}

\caption{Average nearest-neighbors
degree as a function of the vertex degree for the real graphs introduced in Fig. \ref{fig2:cls}.}
\label{fig2:dnn}

\end{figure*}

In Fig. \ref{fig2:cls} we plot $\left<c\right>_d$ vs $d$ for different real networks.
According to this measure, two different classes emerge. On the first class (Math and Route data),
$\left<c\right>_d$ does not exhibit a strong dependency with $d$, except for finite size
effects at the largest degrees. This behavior is typical of random graphs, where the probability
that two neighbors of a vertex are connected by an edge is a constant, and equal two the
probability that any two vertices selected at random are connected. On the contrary, there is
another class where $\left<c\right>_d$ follows an evident decay with increasing the vertex
degree $d$. Thus, in this case, low degree vertices form local sub-graphs that are well
connected. At the same time they are connected to other parts of the graph by high degree  
vertices, having a few edges between the subgraphs they connect but giving a small average
minimum path distance. This picture makes evident the existence of some hierarchy
\cite{pvv01,vpv02a} or modularity \cite{rb02}.

These observations for the clustering coefficient are complemented by another metric related
to the
correlations between vertex degrees. These correlations are quantified by the probability $p(d'|d)$ that a
vertex with degree $d$ has an edge to a vertex with degree $d'$. With the available data a plot of this
magnitude results very noisy and difficult to interpret. Thus in \cite{pvv01} it was suggested to measure
the average degree among the nearest neighbors of a vertex, which is given by
\begin{equation}
\left<d_{nn}\right>_d=\sum_{d'}d'p(d'|d),
\label{dnn}
\end{equation}
and to plot it as a function of the vertex degree $d$. If there are not degree-degree correlations then
the probability that an edge points to a vertex of degree $d'$ is independent of $d$ and 
proportional to $d'p_{d'}$ resulting, after normalization, $p(d'|d)=d'p_{d'}/\left<d\right>$.
Therefore, the plot $\left< d_{nn}\right>_d$ vs. $d$ will be flat and equal to
\begin{equation}
\left<d_{nn}\right>_{\mbox{unco}} = \frac{\left<d^2\right>}{\left<d\right>}.   
\label{dd_rg}
\end{equation}

In Fig. \ref{fig2:dnn} we plot $\left<d_{nn}\right>$ vs $d$ for several real networks. Also in this case
we found the emergence of two different classes of graphs. In one of them the average nearest neighbor
degree exhibits a power law decay with increasing vertex degree. This is a strong evidence of the
existence of disassortative (or negative) correlations, where large degree vertices tend to be connected
with low degree ones and viceversa. On the other hand, for some of the graphs (Math and Route data) an
increasing tendency is
observed denoting the presence of assortative (or positive) correlations, where the edges connect vertices
with similar degrees. The same conclusions are obtained using the Pearson coefficient of the degrees at either ends of an 
edge \cite{n02a,n02e}.
Notice that the subdivision attending either the clustering coefficient or the average nearest-neighbor
degree coincides. 

These observations cover a wide range of networks and are complemented by Refs.
\cite{pvv01,vpv02a,rb02,n02a,n02e}. However, their origin is not yet clear. After some years of intensive
research on complex networks there is not an explanation for the ubiquity of the linear the preferential attachment.
Different models have been proposed but a mechanism is still missing. The lack of a general
principle is extended to these new metrics associated with correlations. In the following sections three
different models that exhibit these properties are studied, emphasizing on the mechanism behind them.
Based on their analysis some general conclusions will be achieved.

\section{Random walk on a net}
\label{sec:rw}

In this section we study the evolution of a graph where we know about new vertices by simply exploring the
graph, with applications to searchable networks such as the citation and WWW graphs. We will focus on
different local mechanisms, where the term local means that we will investigate
evolution rules that involve a vertex and its neighbors. A global approach based on effective
attachment rates can be found in \cite{kr02a}.

There are different ways to obtain information about the documents
(articles, web pages) in these graphs, like looking at directories (citation index,
web crawler), commercial spots, pointed by a friend, or following the references
(citations, hyper-links) that are contained in the documents that we already know. In
the case of the citation graph, we often find new articles from the citation
list of an article that we already know and, later on, we can repeat the process with
these new articles. Moreover, it is known that with a high probability
people know about new web pages by surfing on the WWW.

Two of the major contributions to how people find out about
new web pages are following the hyper-links of other web pages and using search engines \cite{wwws98}. The
first
source can be characterized modeling the WWW ``surfers'' as random walkers on the WWW graph. Let us
assume that the walk starts from a page selected at random and, on each page, with probability
$q_e$ it decides to follow one link on that page or to jump to another random page with probability 
$1-q_e$. Then,
the probability $v_i$ that a page $i$ will be visited is given by
\begin{equation}
v_i = \frac{1-q_e}{N} + q_e\sum_jJ_{ij}\frac{v_j}{d^{ou}_j},
\label{rw1}
\end{equation}
where $J_{ij}$ is the adjacency matrix and $d^{ou}_j$ denotes the vertex out-degree. It is quite
interesting to notice that this probability
of being visited by a random surfer is often used by search engines as a page rank criteria
\cite{hhmn}, as it is the case of the popular Google \cite{bp98}. Hence, the two main sources
through which new pages are visited are characterized by Eq. (\ref{rw1}) and, therefore, the
main properties of the in-degree distribution of the WWW graph should be computed starting on
it. However, up to my knowledge and except from the recursive search model proposed by the author in
Ref. \cite{v00}, no study has been performed in this direction.

In a mean-field approximation one can replace the sum in Eq. (\ref{rw1}) by
$\Theta d^{in}_i$, resulting
\begin{equation}
v_i=\frac{1-q_e}{N}+q_e\Theta d^{in}_i.
\label{eq:walk1}
\end{equation}
where $\Theta$ is the average probability  that a vertex pointing to vertex $i$ is visited and
$d^{in}_i$ is the vertex in-degree. To compute $\Theta$ we
should take into account that the probability that a
vertex $i$ has an in-edge coming from a vertex with out-degree $d^{ou}$ is
$d^{ou}p_{d^{ou}}/\langle d^{ou}\rangle$. This edge will be selected at
random among the $d^{ou}$ out-edges and, therefore, with probability
$1/d^{ou}$. Thus,
\begin{equation}
\Theta = \sum_{d^{ou}} \frac{ d^{ou}p_{d^{ou}} }{ \left<d^{ou}\right> }
\frac{1}{d^{ou}}v_{d^{ou}} = \frac{ \left<v\right>} {\left<d^{ou}\right>}.
\label{eq:walk2}
\end{equation}

In general when we visit new pages we do not create a hyper-link to it. In a first
approximation
this can be modeled introducing a probability $q_v$ that a visited vertex (page) increases its in-degree
by one (a hyper-link is created to it).
Then, when a walk is performed $\left<v\right>N$ vertices are visited and,
therefore, $q_v\left<v\right>N$ edges
are added in
average, resulting
\begin{eqnarray}
\frac{\partial N}{\partial t} &=& \nu_a\ ,
\nonumber\\
\frac{\partial E}{\partial t} &=& \nu_s q_v\left<v\right>N\ ,
\end{eqnarray}
where $E$ is the number of edges, and $\nu_s$ and $\nu_a$ are the number of surfers and the number of
newly
added pages per unit time,
respectively. The integration of these Eqs. yields
\begin{equation}
\langle d^{ou}\rangle=\langle d^{in}\rangle=q_v\left<v\right>N\frac{\nu_s}{\nu_a}.
\label{eq:walk4}
\end{equation}
Thus, from Eqs. (\ref{eq:walk2}) and (\ref{eq:walk4}) we finally obtain
\begin{equation}
\Theta=\frac{\nu_a}{q_v\nu_sN}.
\label{eq:walk5}
\end{equation}

The probability that the in-degree of a vertex of in-degree $d^{(in)}$ increases by one when
a surfer walks on the graph is given by $A(d^{(in)})=q_vv(d^{(in)})$ and, therefore, from
Eqs. (\ref{eq:walk1}) and (\ref{eq:walk5}) it follows that
\begin{equation}
A(d^{(in)})=\frac{1}{N}\left[q_v(1-q_e)+q_e\frac{\nu_a}{\nu_s}d^{(in)}\right].
\label{eq:walk6}
\end{equation}
Notice that the walk on the graph leads to an effective linear preferential attachment.
The degree distribution corresponding to this attachment rate can be easily obtained
using the rate equation approach \cite{krl00,kr01}. Indeed, the number
of vertices $n_{d^{in}}(t)$ with in-degree $d^{in}$ satisfies the rate equation
\begin{equation}
\frac{\partial n_{d^{in}}}{\partial t}=\nu_sA_{d^{in}-1}n_{d^{in}-1}-
\nu_sA_{d^{in}}n_{d^{in}}+\nu_a\delta_{d^{in}0}.
\label{eq:walk7}
\end{equation}
Now we should take into account that the number of vertices on the WWW graph grows
exponentially and, in such a case, $\nu_a\propto N$. Moreover, assuming that each surfer has
its own (or group of) web page (pages) the number of surfers is expected
to be proportional to the number of web pages, {\em i.e.} $\nu_s\propto N$. Thus, 
\begin{equation}
\frac{\nu_s}{\nu_a} = \alpha,
\label{rw2}
\end{equation}
where $\alpha$ is a constant. It is worth noticing that Eq. (\ref{rw2}) is always satisfied for networks with a constant grow
rate, as it may be the case of the citation graph.
If this condition
is satisfied then the in-degree distribution reaches a stationary state and we can write
$n_{d^{in}}(t)=Np_{d^{in}}$, where $p_{d^{in}}$ is
the
stationary probability that a vertex has in-degree $d^{in}$. Substituting this
expression in Eq. (\ref{eq:walk7}) we obtain
\begin{equation}
p_{d^{in}} = \frac{1}{1+a} \frac{ \Gamma[a(\gamma-1)+d^{in}] }{ \Gamma[a(\gamma-1)] } 
\frac{ \Gamma[(1+a)(\gamma-1)+1] }{ \Gamma[(1+a)(\gamma-1)+d^{in}+1] }
\label{eq:walk8a}
\end{equation}
where
\begin{equation}
\gamma = 1+ \frac{1}{q_e},\ \ \ \ a=\alpha q_v(1-q_e)
\label{eq:walk8}
\end{equation}
with the asymptotic behavior for large in-degree 
\begin{equation}
p_{d^{in}}\sim(d^{in})^{-\gamma}.
\label{eq:walk9}
\end{equation}

Hence, the random walk model on a directed graph leads to a power law in-degree distribution,
with an exponent $\gamma\geq2$. Notice that the power law exponent does not depend on $q_v$ and,
therefore, we expect that generalizations of the rule of creating an edge to a visited vertex
would not change this exponent. For instance, one can divide
the vertices in classes in such a way that the edges can be only created among vertices of the
same class, and the resulting power law exponent should be the same. Moreover, the power law exponent does not depend on
$\alpha$. 

We can go beyond the in-degree distribution and compute the clustering coefficient as
a function of the total degree $d=d^{in}+d^{ou}$ of a vertex. For this purpose
we
consider the graph as undirected and compute the number $e_i$ of edges among the
neighbors of a vertex $i$. Since the only dynamics in this model is given by the random
walk it results that
\begin{equation}
\frac{\partial e_i}{\partial t}=q_v\left( q_e\Theta d^{in}_i + q_e v_i \right).
\label{eq:walk10}
\end{equation}
The first term in the right hand side is the probability that a vertex with an out-edge to $i$
is visited and the second the probability that vertex $i$ is visited and the walk follows one of it out-edges to visit an
out-neighbor vertex.
In all cases the visited vertex is
selected with probability $q_v$. Using Eqs. (\ref{eq:walk1}), (\ref{eq:walk5}) and \ref{eq:walk6} and
taking into 
account that $\partial_t d^{in}_i=A(d^{in}_i)$ we can rewrite
(\ref{eq:walk10}) as
\begin{equation}
\frac{\partial e_i}{\partial t} \approx (1+q_e)
\frac{\partial d^{in}_i}{\partial t} ,
\label{eq:walk10a}
\end{equation}
where we have neglected the first term in the right hand side of Eq. (\ref{eq:walk6}).
Integrating this equation with the boundary condition $e(d^{in}=0)=0$ we obtain the clustering
coefficient.
\begin{equation}
\left<c\right>_d=\frac{2e(d)}{d(d-1)}=\frac{2(1+q_e)}{d}+
\frac{2(1+q_e)(1-d^{ou})}{d(d-1)},
\end{equation}
For large $d$ the clustering coefficient scales as
\begin{equation}
\left<c\right>_d\approx\frac{2(1+q_e)}{d}.
\label{cd}
\end{equation}
Thus, we obtain an inverse proportionality between the clustering coefficient and the vertex degree.

\begin{figure}[t]
\begin{center}
\includegraphics[width=3in]{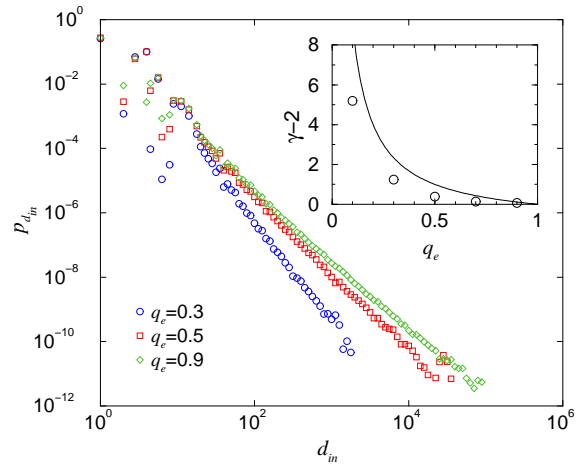}
\end{center}

\caption[In-degree distribution of the random walk model.]{In-degree distribution of the random
walk model for different values
of the probability to continue the walk
$q_e$ and for graph 
size $N=10^6$. In all cases we take average over 100 realizations. The inset shows the exponent
$\gamma$ obtained from the fit to the power law $p_{d^{in}}\sim(d^{in})^{-\gamma}$
(circles) together with the
analytical prediction (continuous line).}

\label{fig3:rw:1}
\end{figure}

\subsubsection{Random walk model}

We now study a particular random walk model by means of numerical simulations and compare its
properties with the analytical
results obtained above. We have made some simplifications in order to reduce the number of parameters and investigate the
influence of the most important parameter $q_e$. The model is defined as follows: {\em Initial
condition}: starting with one
vertex and an empty set of edges, iteratively perform the following rules,
\begin{itemize}

\item{\em Adding}:  A new vertex is created with an edge pointing to one of
the existing vertices, which is selected at random. 

\item{\em Walking}: if an edge is created to a vertex in the network then with
probability $q_e$ an edge is also created to one of its nearest neighbors. When no
edge is created go to the {\em adding} rule.

\end{itemize}

\begin{figure}[t]
\begin{center}
\includegraphics[width=3in]{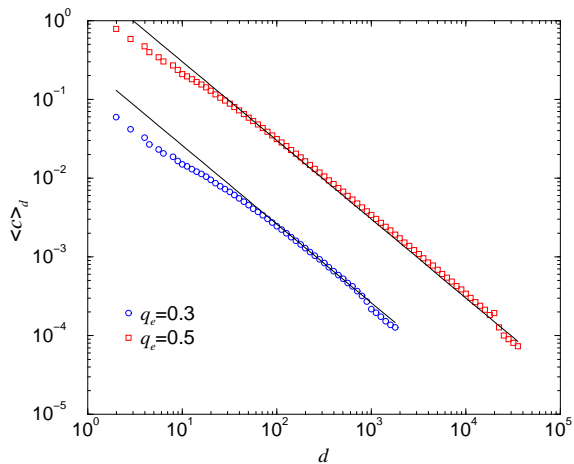}
\end{center}

\caption[Clustering coefficient as a function of vertex degree of the random walk model.]{Clustering coefficient as a
function of vertex degree of the random walk model, for different values of the
probability to continue the walk $q_e$ and for graph size $N=10^6$. In all cases we take average over 100
realizations. The
solid lines correspond with the power law decay $C(d)=2(1+q_e)/d$.}

\label{fig3:rw:2}
\end{figure}

The first simplification is that there is only one ``surfer'' in the network, {\em i.e.} $\nu_s=1$. Second, each time the
``surfer'' decides not to follow one of the edges of the visited vertex it stops, and a new vertex starts a
search from a vertex selected at random. In other words the jump to a random vertex is coupled with the addition of
new vertices resulting $\nu_a=1-q_e$. Finally, each time a vertex is visited an edge is created to it, thus
$q_v=1$. Hence, the in-degree distribution is given by Eq. (\ref{eq:walk8a}) with
\begin{equation}
\gamma = 1 + \frac{1}{q_e},\ \ \ \ a = 1.
\label{par1}
\end{equation}
We have made numerical simulations of this random walk model up to graph sizes $N=10^6$
making average over 100 realizations. In Fig. \ref{fig3:rw:1} we show a log-log plot of the
in-degree distribution for different values of $q_e$. The power law decay for large in-degrees is
evident. The exponent $\gamma$ obtained from the fit to the numerical data is shown in the
inset, together with the predicted dependency in Eq. (\ref{par1}). The analytical values
overestimate the power law exponent but the qualitative picture is the same. For
$q_e\rightarrow0$ the power law exponent is so large that the degree distribution cannot be
distinguished from an exponential distribution. On the contrary, for $q\rightarrow1$ it approaches is
minimum
value $\gamma=2$. We attribute the quantitative disagreement to the mean-field approximation
performed in the step from Eq. (\ref{rw1}) to (\ref{eq:walk1}). On the other hand, the behavior
of the average clustering coefficient with respect to the vertex degree is shown in Fig.
\ref{fig3:rw:2}. In this case the analytical asymptotic behavior in Eq. (\ref{cd}) is in very
good agreement with the numerical data.

\begin{figure}[t]
\begin{center}
\includegraphics[width=3in]{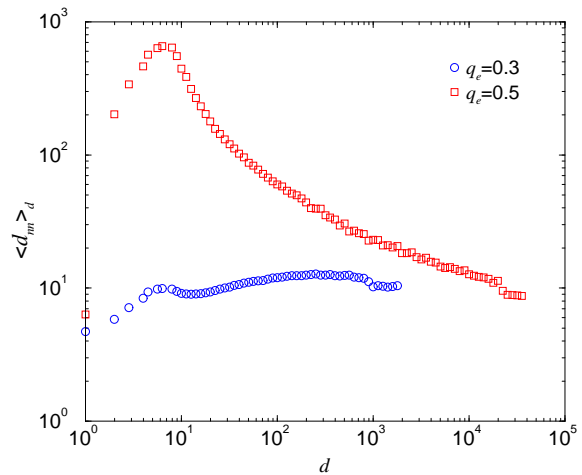}
\end{center}

\caption[Average neighbor degree as a function of vertex degree of the random walk model.]{Average neighbor degree as a
function of vertex degree of the random walk model, for different values of the
probability to continue the walk $q_e$ and for graph 
size $N=10^6$. In all cases we take average over 100 realizations.}

\label{fig3:rw:3}
\end{figure}

We were not able to obtain a prediction for the scaling of the average neighbor degree with the
vertex degree. In this case our analysis relies on numerical simulations. In Fig.
(\ref{fig3:rw:3}) we plot $\left<d_{nn}\right>$ vs. $d$ for two values of $q_e$. For $q_e=0.3$
and for small values of $q_e$ the average neighbor degree does not exhibit a strong dependency
with $d$ and, therefore, the graph appears as uncorrelated. On the contrary, for $q_e=0.5$ and
in general for larger values of $q_e$ it shows a peak around $d=10$ and then decays with
increasing the degree. This decay becomes even faster with increasing $q_e$. We have not found an
explanation for this qualitative change of behavior yet. It is worth noticing that the
experimental data for the WWW yields a $\gamma\approx2.1$, that can be obtained with our model using
$q_e>0.5$. For this value of $q_e$ the model yields negative correlations in agreement with the
real data presented in Sec. \ref{sec:exp}. However, we should take into
account that the above analysis includes the fluctuation properties of the in-degree, while the
statistics of the out-degree was not considered. The last one is irrelevant to determine the in-degree
distribution but has to be taken into account to determine the clustering and degree correlation
properties of the undirected representation of the directed graph. Hence, the results obtained here for
$\left<c\right>_d$ and $\left<d_{nn}\right>_d$ are not conclusive.

\subsubsection{Recursive search model}

In the random walk model one follows only one edge of the visited vertices. However, one may consider an exhaustive 
search
following all the edges recursively \cite{v00}. The main idea of a recursive search is thus to be
connected to one vertex of the
network
and any time we get in contact with a new vertex we follow all its edges, exploring in this
way a larger part of the network. This can be modeled modifying the walking rule as follows,
\begin{itemize}

\item{\em Walking}: if an edge is created to a vertex in the network then with
probability $q_e$ an edge is also created to each of its nearest neighbors. When no
edge is created go to the {\em adding} rule.

\end{itemize}
As for the previous model we have $\nu_s=1$, $\nu_a=1-q_e$ but $A(d^{in})$ is not given by
Eq. (\ref{eq:walk6}). The form of $A(d^{in})$, and consequently the in-degree distribution, is determined below for
two limiting cases.

$q_e=0$: In this case only the {\em adding} rule is performed, hence $A(d^{in})=1/N$
independent of $d^{in}$. The fact that $A(d^{in})$ scales as $N^{-1}$ carries as a consequence
that $n_{d^{in}}(N)=Np_{d^{in}}$ is the stationary solution of eq. (\ref{eq:walk7}), where $p_{d^{in}}$ is
the stationary probability to find a vertex with in-degree $d^{in}$. Substituting this
expression in Eq. (\ref{eq:walk7}) one obtains 
\begin{equation}
p_{d^{in}}=2^{-(d^{in}+1)}.
\label{eq:rc2}
\end{equation}
$q_e=1$: Also for this limiting case the in-degree distribution can be computed exactly. Let us determine
$A(d^{in})$ using the following fact. Any vertex $i$ with
in-degree $d^{in}_i$ has $d^{in}_i$ vertices with an edge to it, which will be denoted by $x_j$
($j=1,2,\ldots,d^{in}_i$). At the same time each of these $x_j$ vertices may have other
vertices with an edge to it. The following result holds: any vertex with an edge to any of
the vertices $x_j$ has also an edge to $i$. The proof is straightforward, if when a vertex
is added it creates an edge to any of the vertices $x_j$ then with probability $q_e=1$ it
creates an edge to all the nearest neighbors of $x_j$, among which vertex $i$ is
contained; end of the proof. Hence, the probability that when a vertex is added it
creates an edge to vertex $i$ is just the probability $(1+d^{in}_i)/N$ that the first edge
is connected to $i$ or to any of the $d^{in}_i$ vertices with an edge to $i$, {\em i.e.}
$A(d^{in})=(1+d^{in})/N$. As for $q_e=0$ $A(d^{in})$ scales as $1/N$ and, therefore, the
stationary solution is of the form $n_{d^{in}}(N)=Np_{d^{in}}$. Then from Eq. (\ref{eq:walk7}) it
follows that 
\begin{equation}
p_{d^{in}}=\frac{1}{(d^{in}+1)(d^{in}+2)}.
\label{eq:rc3}  
\end{equation}
Notice that also in this case, although it is not implicitly assumed, there is a
preferential attachment leading to the power-law decay for large in-degrees $p_{d^{in}}\sim(d^{in})^{-2}$.

%
%

\begin{figure}[t]
\begin{center}
\includegraphics[width=3in]{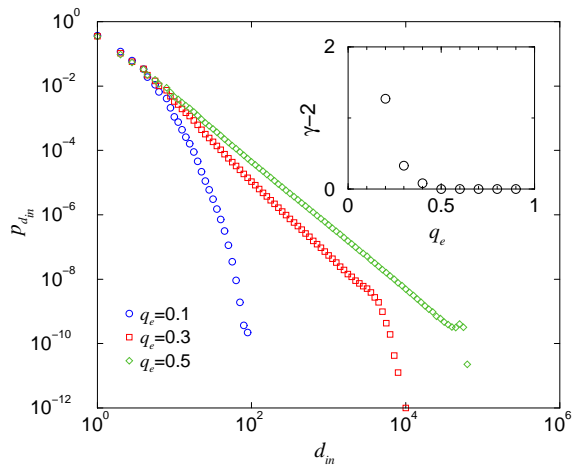}
\end{center}
\caption[Log-log plot of the in-degree distribution of the recursive search model.]{Log-log plot of the in-degree
distribution of the recursive search model for different values of $q_e$. The inset shows the exponent $\gamma$ 
obtained from the power law fit $p_{d^{in}}\sim (d_{in})^{-\gamma}$ to the numerical data.}
\label{fig3:rw:rc3} 
\end{figure}

The limiting cases $q_e=0$ and $q_e=1$ are described by in-degree distributions which
are qualitative different.  For $q_e=0$ the distribution is exponential with a finite
average in-degree. On the contrary, for $q_e=1$, the distribution follows a power law
decay $p_{d^{in}}\sim {d^{in}}^{-\gamma}$ for large $d^{in}$, with $\gamma=2$. This power law decay
goes up to the largest possible degree $d^{in}\sim N^{1/(\gamma-1)}\sim N$ while $p_{d^{in}}=0$ for $d^{in}\geq N$.
Hence, for $q_e=1$ and large $N$ the average in-degree scale as
\begin{equation}
\langle d^{in}\rangle(N) = \langle d^{ou}\rangle(N) = a+\ln N,
\label{eq:rc8}
\end{equation}
where $a$ is independent of $N$ and clearly $\langle d^{in}\rangle$ diverges in the thermodynamic
(large network sizes) limit. In a mean-field approximation one can neglect the existence of loops in the network and, in
such a
case, the ``walking'' rule will take place on a tree. Each vertex on the tree will have on
average $\langle d^{ou}\rangle(N)$ sons, which is just the average out-degree after $N$ vertices
have been added. Moreover, if a vertex is visited then each of its sons will be visited with
probability $q_e$. Hence, when the vertex $N+1$ is added, its average out-degree $\langle
d^{ou}\rangle(N+1)$ will be given by the average number of vertices visited during the walk, {\em
i.e.}
\begin{eqnarray}
\langle d^{ou}\rangle(N+1)
&=& 1+q_e\langle d^{ou}\rangle(N)+[q_e\langle d^{ou}\rangle(N)]^2+
\ldots
\nonumber\\
&=& \frac{1}{1-q_e\langle d^{ou}\rangle(N)}.
\label{eq:rc8a}
\end{eqnarray}
If there is a stationary state then $\langle d^{ou}\rangle(N+1)=\langle d^{ou}\rangle(N)=\langle
d^{ou}\rangle$. In this case Eq. (\ref{eq:rc8a}) yields two solutions. One of them diverges when
$q_e\rightarrow0$, which is not admissible since $\langle d^{ou}\rangle=1$ for $q_e=0$. The
other solution reads
\begin{equation}
\left<d^{ou}\right> = \langle d^{in}\rangle=\frac{1-\sqrt{1-4q_e}}{2q_e}.
\label{eq:rc8b}
\end{equation}
This solution is valid for $q_e\leq q_c=1/4$ and, therefore, the average out degree does not
converge to an stationary value when $q_e>q_c$. In this last region the average out degree
increases logarithmically with $N$, as in the extreme case $q_e=1$ (see
Eq. (\ref{eq:rc8})). Now, $\left<d^{in}\right> = \left<d^{ou}\right>$ and both
approach a stationary state for any $\gamma>2$ and diverge otherwise. We then expect that the in-degree
distribution has a
power law exponent $\gamma>2$ for $q_e<q_c$ and $\gamma\leq2$ for $q_e>q_c$. Moreover, taking
into account that the fastest
divergence is obtained for $q_e=1$, where $\gamma=2$, we conclude that for $q_e>q_c$ the power law
exponent is constant
and equal to $\gamma=2$.

To investigate the behavior for $0<q_e<1$ and the existence of a non trivial threshold $q_c$ as predicted by the
mean-field approach, we have made numerical simulations of the recursive search model for different values of $q_e$ up to
graph sizes $N=10^5$. For each value of $q_e$ the in-degree distribution was averaged over 100 runs of the algorithm.
The resulting in-degree distribution is shown in Fig. (\ref{fig3:rw:rc3}). For $q_e=0.1$ the decay for
large in-degrees is
very fast, and can be fitted by a power law decay with a very large exponent or equivalently by an exponential decay. On
the contrary, for larger $q_e$ the exponent becomes smaller and the power law behavior becomes more evident. Finally, for
$q_e\geq q_c=0.5\pm0.1$ the exponent becomes independent of $q_e$ and equals $\gamma=2$, in
agreement with the mean-field
prediction. However, the numerical threshold is two times the value obtained from Eq. (\ref{eq:rc8b}).

In ordinary critical phenomena the absence of any typical length scale takes place
at the critical point, which is observed at a precise value of the order
parameter. For the present model, however, the absence of a characteristic
in-degree is not only manifested at a precise value of $q_e$ but in the whole
interval $q_c\leq q_e\leq1$.  These features are very similar to those observed in
some sandpile models \cite{td97,vs99}, the paradigm of self-organized critical systems
\cite{btw87,btw88}. As in these models \cite{vz97,vz98}, there is a time scale
separation between the addition of new vertices and their ``walk'' through the network.
In the thermodynamic limit ($N\rightarrow\infty$) the phase diagram of the model is
divided in a sub-critical ($0\leq q_e<q_c$) and a critical region ($q_c\leq
q_e\leq1$), where the power law exponent does not depend on the
control parameter.  Hence, the results presented here suggest that for $q_c\leq
q_e\leq1$ the present model is in a self-organized critical state.

\section{Connecting nearest-neighbors}
\label{sec:cnn}

In social graphs it is more probable that two vertices with a common neighbor get connected than two vertices chosen at
random \cite{n01b}. Clearly this property leads to a large average clustering coefficient since it increases the number
of connections between the neighbors of a vertex, as it has been already observed in a model proposed by Davidsen, Ebel
and Bornholdt (DEB) \cite{deb01a}.  The basic assumption of their model is that the evolution of social connections is
mainly determined by the creation of new relations between pairs of individuals with a common friend. Moreover, a similar
mechanism was considered by Holme and Kim \cite{hk02} and by Gin {\em et al} \cite{jgn01} to introduce an appreciable
clustering coefficient in preferential attachment models.

The study of these models has been mainly performed by numerical simulations. 
A deeper analytical understanding can be obtained by introducing the concept of potential edge.
We will say that a pair of vertices is connected by a {\em potential edge} if
\begin{enumerate}
\item they are not connected by an edge and 
\item they have at least one common neighbor. 
\end{enumerate}
Notice that while this concept have been implicitly 
considered in previous works its mathematical description will be introduced here for the first time. 

The graph dynamics will be defined by the
transition rates between the three possible states of a pair of vertices: disconnected ($s$), connected by
a potential edge ($p$) or by an edge ($e$). Let $d^*_i$ be the number of potential edges incident to
vertex $i$, potential degree to abbreviate. We can write the rate equations for the evolution of the
number of vertices with degree $d$ and potential degree $d^*$. Instead we will use the continuum approach
\cite{baj00b,dms00}. In this case we neglect fluctuations
and write mean-field equations for the evolution of $d_i$ and $d^*_i$,
\begin{eqnarray}
\frac{\partial d_i}{\partial N} &=& \nu_{s\rightarrow e}\hat{d}_i + \nu_{p\rightarrow e} d^*_i 
- (\nu_{e\rightarrow s} + \nu_{e\rightarrow p}) d_i,
\nonumber\\
\frac{\partial d^*_i}{\partial N} &=& \nu_{s\rightarrow p}\hat{d}_i + \nu_{e\rightarrow p} d_i 
- (\nu_{p\rightarrow s} + \nu_{p\rightarrow e}) d^*_i\ ,
\nonumber\\
\hat{d}_i &=& N-d_i-d^*_i\ .
\label{RE1}
\end{eqnarray}
$\nu_{x\rightarrow y}$ is the transition rate from state $x$ to state $y$ per unit of $N$ and
$\hat{d}_i$ is the number of remaining neighbors, that are not connected by a potential edge nor by an
edge to vertex $i$. 

The creation (deletion) of a potential edge incident to a vertex is associated with the
creation (deletion) of an edge incident to one of its neighbors. For instance, if a new vertex $i$ is
connected to an existing vertex $j$ then a potential edge is created between $i$ and all neighbors of
$j$. Hence
\begin{eqnarray}
\nu_{s\rightarrow p} = \nu_{s\rightarrow e} d_i,
\nonumber\\
\nu_{p\rightarrow s} = \nu_{e\rightarrow s} d_i.
\label{p_s}
\end{eqnarray}
These equalities are at the core of the connecting nearest-neighbors model.

In the following we will neglect any process where an edge is deleted, {\em i.e.} 
\begin{equation}
\nu_{e\rightarrow s} =0.
\label{assumption1}
\end{equation}
This assumption may seem too crude for some social networks where it is known that social relations can be lost
but it is realistic in many other cases. For instance, in the network of scientific collaborations two
scientists are said to be connected if they have co-authored a paper. It is clear that this connection
cannot be lost in time because the fact that they have written a paper together cannot be changed.
In general, if the connection between two vertices is given by the occurrence of certain event
(co-authoring a paper, being in the cast of a the same film, having a sexual relation) in the past history
then this connection cannot be lost and, therefore, our approximation holds. 

Another crucial assumption is
related to the fact that the transition from potential edge to an edge has a higher probability of
occurrence than the transition from disconnected to an edge. In fact, the connection of two disconnected
vertices without a common neighbor is a process that models the creation of a social relation between two
social entities chosen at random. We thus assume
\begin{equation}
\nu_{s\rightarrow e} = \frac{\mu_0}{N^2}.
\label{mu0}
\end{equation}
On the other hand, the creation of an edge between two vertices with a common neighbor, that is
with a potential edge
between them, models the creation of a social relation between two ``friends'' of a social entity. In this case we assume
\begin{equation}
\nu_{p\rightarrow e} = \frac{\mu_1}{N}.
\label{mu1}
\end{equation}

\begin{figure}[t]
\begin{center}
\includegraphics[width=3in]{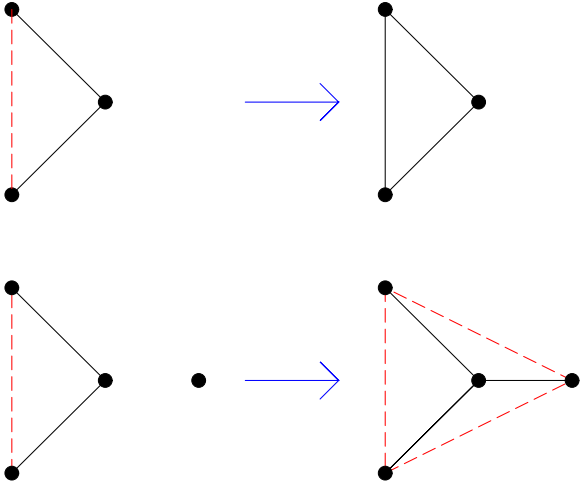}
\end{center}

\caption[Schematic representation of the two evolution rules of the connecting nearest-neighbors model.]{Schematic
representation of the two evolution rules of the connecting nearest-neighbors model. Top: with probability $u$ a
potential edge (dashed line) becomes an edge (continuum lines). Bottom: with probability $1-u$ a new vertex is added to
the graph (disconnected vertex in the left), then it is connected with an edge to a vertex selected at random and by
potential edges to its neighbors (right).}

\label{fig3:c2n:draw:1}
\end{figure}

Under these approximations the system of equations (\ref{RE1}) is reduced to
\begin{eqnarray}
N\frac{\partial d_i}{\partial N} = \mu_0 + \mu_1 d^*_i,
\nonumber\\
N\frac{\partial d^*_i}{\partial N} = \mu_0 d_i - \mu_1 d^*_i,
\label{RE2}
\end{eqnarray}
Hence, the existence of a linear preferential attachment (the growth rates of $d_i$ and $d^*_i$ are linear in
themselves) in this class of models becomes evident
with the
introduction of the concept of potential edges. Thus, a power-law degree distribution is expected.
This system of differential equations is linear and, therefore, can be easily integrated
resulting that, for $N\gg N_i$, 
\begin{equation}
d_i(N) = d_0\left(\frac{N}{N_i}\right)^\beta,\ \ \ \ 
d^*_i(N) = d^*_0\left(\frac{N}{N_i}\right)^\beta,
\label{d_N}
\end{equation}
where $N_i$ is the size of the graph when vertex $i$ was added to it and
\begin{equation}
\beta = \frac{\mu_1}{2}\left(-1+\sqrt{1+4\frac{\mu_0}{\mu_1}}\right).
\label{beta}
\end{equation}
Now, if the vertices are added at a
constant rate then $P(N_i=N)=1/N$ yielding
\begin{eqnarray}
P(d_i>d) &=& P\left[ d_0 \left(\frac{N}{N_i}\right)^\beta > d \right]
\nonumber\\
&=& \int_0^N \frac{dN_i}{N} \Theta\left[d_0 \left(\frac{N}{N_i}\right)^\beta - d \right],
\end{eqnarray}
Consequently,
\begin{equation}
p_d = \frac{\partial P(d_i>d)}{\partial d}\sim d^{-\gamma}\ .
\end{equation}
with
\begin{equation}
\gamma = 1 + \frac{1}{\beta}.
\label{gamma1}
\end{equation}
Notice that the main ingredient leading to this power law behavior is given by Eq. (\ref{p_s}). On the contrary, if
$\nu_{s\rightarrow p}$ would be independent of the vertex degree an exponential decay would be obtained.

\begin{figure}[t]
\begin{center}
\includegraphics[width=3in]{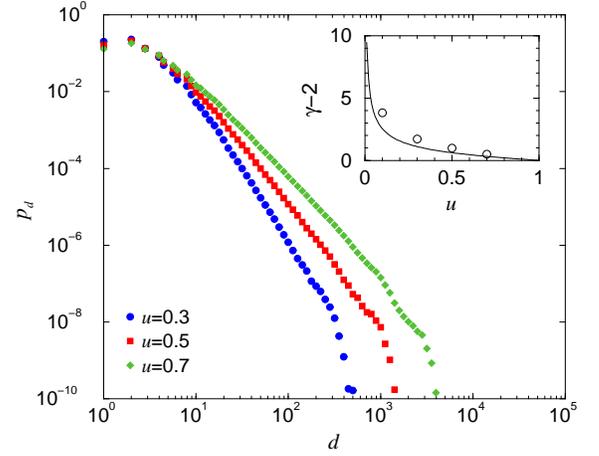}
\end{center}

\caption[Degree distribution of the connecting nearest neighbors model.]{Degree distribution of
the connecting nearest neighbors model for different values of the addition rate $u$, graph size
$N=10^6$ and average over 100 realizations. The inset shows the exponent $\gamma$ obtained from
the fit to the power law $p_d=ad^{-\gamma}$ (circles) together with the analytical prediction
(continuous line).}

\label{fig3:c2n:1}
\end{figure}

We can also compute the clustering coefficient as a function of the vertex
degree. The main contribution to the evolution of $e_i$, the number of edges among the neighbors of vertex $i$, is given
by the transition {\em potential edge} $\longrightarrow$ {\em edge}. In fact, if the potential
edge connecting a vertex $i$ to another
vertex $j$, with common neighbor $k$, becomes an edge then vertex $i$ gains one neighbor (vertex $j$) and a
new edge among
its neighbors (that connecting $j$ and $k$). Neglecting other contributions we have
\begin{equation}
\frac{\partial e_i}{\partial N} = \nu_{p\rightarrow e} d^*_i = \mu_1 \frac{d^*_i}{N}.
\end{equation}
Integrating this equation using Eq. (\ref{d_N}) it results that
\begin{equation}
\left<c\right>_d = \frac{2e(d)}{d(d-1)}\approx\frac{2\mu_1}{d}.
\label{eq:neighbor6}
\end{equation}
Thus, once again we obtain the inverse proportionality between $\left<c\right>_d$ and vertex degree $d$, in this case due
to the conversion of potential edges between vertices with a common neighbor into edges.

\begin{figure}[t]
\begin{center}
\includegraphics[width=3in]{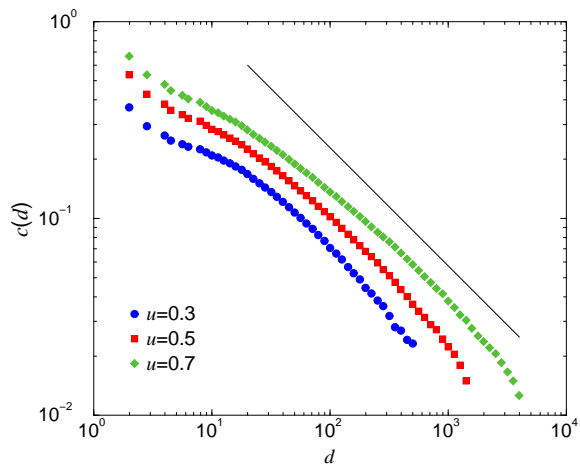}
\end{center}

\caption[Clustering coefficient as a function of vertex degree of the connecting nearest neighbors model.]{Clustering
coefficient as a function of vertex degree of the connecting nearest neighbors model for different
values of the addition rate $u$, graph size $N=10^6$ and average over 100 realizations. The solid line is a power law
decay with exponent 0.6.}

\label{fig3:c2n:2}
\end{figure}

\subsubsection{Connecting nearest-neighbors model}

To check these results we have made numerical simulations of a variant of the DEB model.
Starting with a single vertex and an empty
set of edges iteratively perform the following rules: 
\begin{itemize}

\item With probability $1-u$ introduce a new vertex in the graph, create an edge from the new vertex to a vertex $j$ 
selected at random,
(implying the creation of a potential edge between the new vertex and all the neighbors of
$j$).

\item With probability $u$ convert one potential edge selected at random into an edge.

\end{itemize}
A schematic representation of these rules is shown in Fig. \ref{fig3:c2n:draw:1}. Actually, in the
DEB model the number of vertices is fixed
and each time a
new vertex is added one vertex is removed from the graph. We consider the growing variant because in this
case it is easier
to determine some properties analytically. For very large $N$ we expect that both variants have the same qualitative
behavior. 

These evolution rules fit into the equations written above after setting
\begin{equation}
\mu_0 = 1,\ \ \ \ \mu_1 = \frac{u}{1-u}.
\label{mu}
\end{equation}
Thus, from Eqs. (\ref{beta}) and (\ref{gamma1}) it follows that
\begin{equation}
\gamma(u) = 1 + \frac{2(1-u)}{u}\left(-1+\sqrt{1+4\frac{1-u}{u}}\right)^{-1},
\label{gamma_u}
\end{equation}
with the limiting cases
\begin{equation}
\gamma(0) = \infty,\ \ \ \ \gamma(1) = 2.
\end{equation}
Thus, the power law exponent $\gamma$ takes its minimum value when $u\rightarrow1$ corresponding to a low
rate of addition of vertices and it grows with decreasing $u$ corresponding to higher rates of vertex
addition. In Fig. \ref{fig3:c2n:1} we plot the degree distribution as obtained from numerical simulations.
For intermediate degrees it exhibits a power law decay $p_d\sim d^{-\gamma}$. The value of $\gamma$
obtained from the fit to the numerical data is shown in the inset, together with the analytical curve
given
by Eq. (\ref{gamma_u}). The quantitative disagreement tell us that the mean-field Eq. (\ref{RE1}) give us
the right qualitative description but fluctuations should be considered to obtain a precise
agreement with the numerical data.

\begin{figure}[t]
\begin{center}
\includegraphics[width=3in]{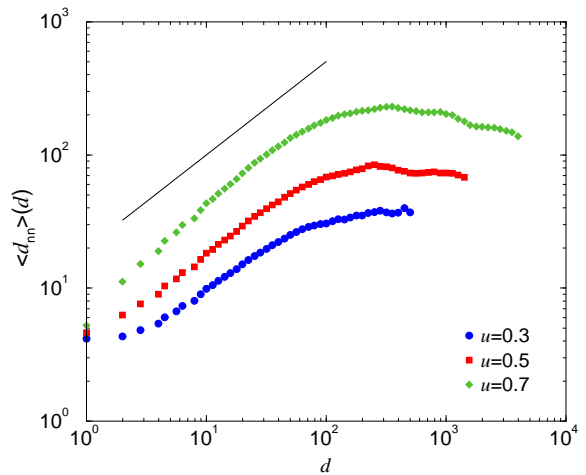}
\end{center}

\caption[Average degree among the neighbors of a vertex with degree $d$ of the connecting nearest-neighbors
model.]{Average degree among the neighbors of a vertex with degree $d$ of the connecting nearest neighbors model for
different values of the addition rate $u$, graph size
$N=10^6$ and average over 100 realizations. The solid line is a power law growth with exponent 0.6.}

\label{fig3:c2n:3}
\end{figure}

In Fig. \ref{fig3:c2n:2} we plot the clustering coefficient as a function of the vertex degree. It follows
a power law decay for large degrees but with an exponent smaller than 1. 
On the other hand, the average neighbor degree as a function of the vertex degree is shown in Fig.
\ref{fig3:c2n:3}. It increases with increasing $d$, {\em i.e.} the graphs generated using this model
exhibit positive degree correlations. This result is in very good agreement with the observations made for
social graphs that are also characterized by positive degree correlations. Hence, the connecting
nearest-neighbors mechanism generates many of the topological properties of social networks, including
power law degree distributions and positive correlations.

\section{Duplication divergence}
\label{sec:copy}

The evolution of some real graphs is given by a replication or partial replication of its local structure.
An example is the genome that evolves, among other mechanisms, through single gene or full genome
duplications \cite{o70} and mutations that lead to the differentiation of the duplicate genes. The
evolution of the genome can be translated into the evolution of the protein-protein interaction network
where each vertex represents the protein expressed by a gene. After gene duplication both expressed
proteins will have the same interactions. This corresponds to the addition of a new vertex in the network
with edges pointing to the neighbors of its ancestor. In addition positive and negative mutations can be
modeled by the creation and lost, respectively, of the edges leading to the divergence of the duplicates
\cite{w01,vfmv01,spsk01a}. The duplication mechanism has been also considered in the evolution of other
biological networks \cite{sk00}. Moreover, another example is given by the WWW where new web pages may be
created making a copy or a partial copy of the hyperlinks present in other web pages \cite{kkrrt99}. In
this case the duplication represents the copying process and the divergence the deletion or addition of
hyperlinks in the duplicated pages.

In a first approximation we will
assume that the processes of duplication and divergence are not coupled but take place
independently one of the other. Moreover, we will also assume that the creation and deletion of
edges take place at random and that they are independent of the degree of the vertices at the
edge ends, or any other topological property. Under these approximations, the evolution of the
degree of a vertex (the number of interacting partners) is given by 
\begin{equation}
\frac{\partial d_i}{\partial N} = \nu_D d_i + \nu_C (N-d_i) - \nu_L d_i,
\label{ddr0}
\end{equation}
where $\nu_D$, $\nu_C$, and $\nu_L$ are the rates per unit of vertex added of duplications, edge creation
and edge
lost, respectively. By definition, each duplication implies the addition of a new vertex and, therefore,
\begin{equation}
\nu_D = \frac{1}{N}.
\label{ddr2}
\end{equation}
We will further assume that
\begin{equation}
\nu_C = \frac{\mu_0}{N},\ \ \ \ \nu_L = \frac{\mu_1}{N}
\label{ddr3}
\end{equation}
otherwise the stationary graph will be empty or fully connected, both being unreal. Notice that $\mu_0$
and $\mu_1$ are new parameters with no relation to those introduced in the previous section.
Then,
substituting
Eqs. (\ref{ddr2}) and (\ref{ddr3}) into Eq. (\ref{ddr0}) we obtain
\begin{equation}
N\frac{\partial d_i}{\partial N} = \mu_0 + (1-\mu_1) d_i.
\label{ddr1}
\end{equation}
The linear dependency of the growth rate with $d_i$ evidences once again the existence of an effective
linear preferential attachment. The integration of this equation yields
\begin{equation}
d_i(N) = \left( d_i(N_i) + \frac{\mu_0}{1-\mu_1} \right) 
\left(\frac{N}{N_i}\right)^\beta - \frac{\mu_0}{1-\mu_1},
\label{dd_d_N}
\end{equation}
where $N_i$ and $d_i(N_i)$ are the graph size and degree of vertex $i$ when vertex $i$ was added to the graph, and
\begin{equation}
\beta = 1-\mu_1.
\end{equation}
Here we have implicitly assumed that
\begin{equation}
\mu_1<1,
\label{rl}
\end{equation}
otherwise the stationary state will be an empty graph.

From Eq. (\ref{dd_d_N}) it follows that
\begin{eqnarray}
P(d_i>d) &=& P\left[ \left( d_i(N_i) + \frac{\mu_0}{1-\mu_1} \right) \left(\frac{N}{N_i}\right)^\beta
\right.
\nonumber\\
&-& \left. \frac{\mu_0}{1-\mu_1}> d \right].
\label{dd_cum1}
\end{eqnarray}
This probability should be computed taking into account that both $N_i$ and $d_i(N_i)$ are random
variables. If the duplications take place at a constant rate then the probability density of $N_i$ is
given by $P(N_i=N)=1/N$. Moreover, the probability that a vertex has degree $d_i(N_i)$
when it is introduced is just the probability that its ancestor has this degree. If the graph is in a
stationary state then $P[d_i(N_i)=d]=p_d$, is just the degree distribution. Hence
\begin{eqnarray}
P(d_i>d) &=& \sum_{d'}p_{d'}\int_1^N \frac{dN_i}{N}
\Theta\left[ \left( d' + \frac{\mu_0}{1-\mu_1} \right) \left(\frac{N}{N_i}\right)^\beta\right. 
\nonumber\\
&-& \left. \frac{\mu_0}{1-\mu_1} > d \right].
\end{eqnarray}
For $N\gg1$ we finally obtain
\begin{equation}
p_d = \frac{\partial P(d_i>d)}{\partial d}\sim \left(\frac{\mu_0}{1-\mu_1}+d\right)^{-\gamma},
\end{equation}
with
\begin{equation}
\gamma = 1 + \frac{1}{1-\mu_1}.
\label{dd_gamma}
\end{equation}
The origin of this power law degree distribution is determined by the second term in the right
hand side of
Eq. (\ref{ddr1}), associated with the vertex duplications and subsequent edge lost. These are local mechanisms and, as in
the models describe before, they lead to an effective preferential attachment manifested as a power law degree
distribution.

The next step is thus to investigate if the duplication-divergence model satisfies the inverse
proportionality between the
average clustering coefficient and vertex degree. If the creation of new interactions takes place at random, {\em i.e.}
they appear between randomly chosen vertices, then the average clustering coefficient will be negligible for large graph
sizes $N$. There is however one source of new interactions giving an appreciable contribution. In the
duplication process, if the ancestor is a self-interacting protein then the ancestor and the duplicate may
have an interaction among them \cite{w01}. Let us assume that this happens with a probability $q_v$. Thus, if a neighbor
of a vertex $i$ is duplicated it will gain a new neighbor (the copy) and with probability $q_v$ an edge between its
neighbors (that between the copy and its ancestor) and therefore
\begin{equation}
\frac{\partial e_i}{\partial t}\approx
q_v\frac{\partial d_i}{\partial t}.
\label{eq:copy4}
\end{equation}
where we have neglected any other process leading to new interactions and edge lost.
The integration of this equation yields
\begin{equation}
\left<c\right>_d=\frac{2e(d)}{d(d-1)}\approx\frac{2q_v}{d}.
\label{eq:copy5}
\end{equation}
Hence, under these assumptions we obtain the inverse proportionality behavior. The inclusion of the edge lost
may change this result. We do not have any analytical proof but since this process contributes to the lost of triangles
and
it has a higher impact in high degree vertices then we expect that $\left<c\right>_d$ would decay faster
than $d^{-1}$.

\begin{figure}[t]
\begin{center}
\includegraphics[width=2in]{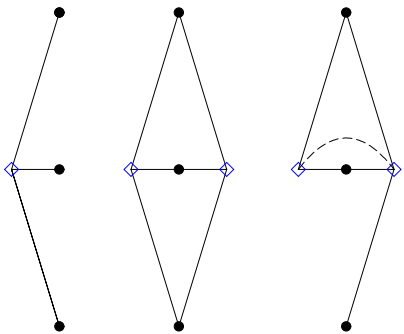}
\end{center}

\caption[Schematic representation of the coupled duplication-divergence model evolution
rules.]{Schematic representation of the coupled duplication-divergence model evolution rules.
Left and middle: A vertex ($\diamondsuit$) is being duplicated. Right: The divergence of the
duplicates is
manifested as a coupled lost of interactions, where the coupling is given by the restriction
that for each neighbor ($\bullet$) at least one of the duplicates should preserve an edge to it.
Moreover, due to the existence of self-interactions, a new edge can be created between the
duplicates (dashed line).}

\label{fig3:dd:draw:1}
\end{figure}

\subsubsection{Coupled duplication-divergence model}

In some practical cases the processes of duplication and divergence cannot be decoupled. For instance, the
protein-protein interaction network has a
functional role in the organism and, therefore, the lost of certain interactions can result in the death of the
corresponding organism. According to the classical model \cite{o70} after duplication the duplicate genes have fully
overlapping functions. Later on, one of the copies may either become nonfunctional due to degenerative mutations or it can
acquire a novel beneficial function and become preserved by natural selection. In a more recent framework
\cite{flp99,lf00} it is proposed that both duplicate genes are subject to degenerative mutations loosing some functions
but jointly retaining the full set of functions present in the ancestral gene. To investigate the influence of the
coupling between duplication and divergence we consider the following model introduced in
Ref. \cite{vfmv01}: At each time step a vertex is
added according to the following rules 
\begin{itemize}

\item{\em Duplication}: a vertex $i$ is selected at random. A new vertex
$i\prime$ with an edge to all the neighbors of $i$ is created.
With probability $q_v$ an edge between $i$ and $i\prime$ is established
(self-interacting proteins).

\item{\em Divergence}: for each of the vertices $j$ connected to $i$ and  $i\prime$
we choose randomly one of the two edges $(i,j)$ or $(i\prime,j)$ and remove
it with probability $1-q_e$.

\end{itemize}
A schematic representation of this rules is shown in Fig. \ref{fig3:dd:draw:1}. A similar model with an
asymmetric divergence has been introduced in Ref. \cite{spsk01a}.
For practical purposes the
algorithm starts with two connected vertices and we repeat the duplication-divergence rules $N$ times. Since
genome evolution analysis \cite{w01,hb98} supports the idea that the divergence of duplicate genes takes
place shortly after the duplication, we can assume that the divergence process always occurs before any
new duplication takes place; i.e., there is a time scale separation between duplication and
mutation rates. This allows us to consider the number of vertices in the network, $N$, as a measure of
time (in arbitrary units). It is worth remarking that the algorithm does not include the creation of new
edges, {\em i.e.} the developing of new interactions between gene products, other than those due to
self-interactions. However, we have tested that the
introduction in the coupled duplication-divergence algorithm of a probability to develop new random
connections does not change the network topology substantially.

\begin{figure}[t]

\begin{center}
\includegraphics[width=3in]{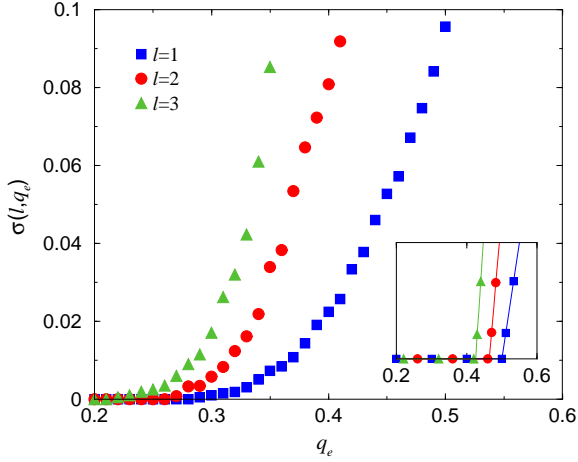}
\end{center}

\caption[Multi-fractal scaling in the coupled duplication-divergence model.]{The exponent $\sigma_l(q_e)$
as a function of $q_e$ for different
values of $l$. The symbols were obtained from numerical simulations of
the model. The moments $\left< d^l\right>$ were computed as a function
of $N$ in networks with size ranging from $N=10^3$ to $N=10^6$.
The exponents $\sigma_l(q)$ are obtained from the power
law fit of the plot $\left< d^l\right>$ vs. $N$. In the inset we
show the corresponding mean-field behavior, as obtained from Eq. (\ref{eq:copy3}),
which is in qualitative agreement with the numerical results.}

\label{fig3:dd1}

\end{figure}

\begin{figure}[t]
\begin{center}
\includegraphics[width=3in]{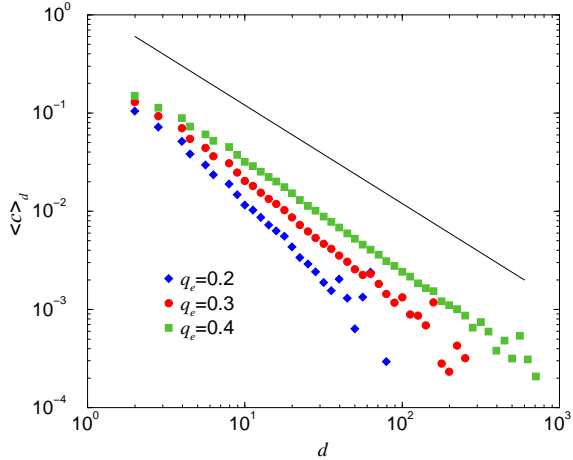}
\end{center}

\caption[Clustering coefficient as a function of vertex degree of the coupled duplication-divergence
model.]{Clustering
coefficient as a function of vertex degree of the coupled duplication-divergence model for different
values of $q_e$, graph size $N=10^6$ and average over 100 realizations. The solid line is a power law 
decay with exponent 1.}

\label{fig3:dd:2}

\end{figure}

\begin{figure}[t]
\begin{center}
\includegraphics[width=3in]{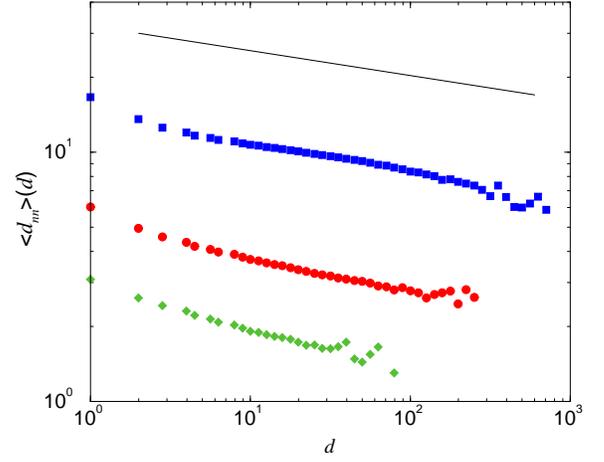}
\end{center}

\caption[Average degree among the neighbors of a vertex with degree $d$ of the coupled
duplication-divergence
model.]{Average degree among the neighbors of a vertex with degree $d$ of the coupled
duplication-divergence model for
different values of $q_e$, graph size
$N=10^6$ and average over 100 realizations. The solid line is a power law decay with exponent 0.1.}

\label{fig3:dd:3}
\end{figure}

In order to provide a general analytical understanding of the  model, we use a mean-field approach for the
moments
distribution behavior.  Let $\left< d\right>(N)$ be the average degree of the network with $N$ vertices. After a
duplication event $N\to N+1$ we have that the average degree is given by
\begin{equation}
\left< d\right>(N+1)=\frac{N\left< d\right>(N) + 2q_v+(2q_e-1)
\left< d\right>(N)}{N+1}.
\label{eq:1}
\end{equation}
On average, the gain will be proportional to $2q_v$ because of the interaction between duplicates, and to
2$\left< d\right>(N)$ because of duplication, and a loss proportional to $2(1-q_e) \left< d\right>(N)$ due
to the divergence process.  For large $N$, taking the continuum limit, we obtain a differential equation
for $\left<d\right>$. For $q_e<1/2$, $\left< d\right>$ grows with $N$ but saturates to the stationary
value
$\left<d\right>=2q_v/(1-2q_e)+{\cal O}(N^{2q_e-1})$, On the contrary, for $q_e>1/2$, $\left<d\right>$
grows with $N$ as $N^{2q_e-1}$. At $q_e=q_1=1/2$ there is a dramatic change of behavior in the large scale
degree properties.  Analogous equations can be written for higher order moments $\left<d^l\right>$. Using
a rate equations approach similar to that considered in Ref.~\cite{kkkr02} it is obtained that
\begin{equation}
\frac{\partial n_d}{\partial N}=A_{d^-1}n_{d^-1}-
A_dn_d - \frac{n_d}{N} + 2q_vG_{d-1} + 2(1-q_v)G_d,
\label{eq:copy2}
\end{equation}
where
\begin{equation} 
A(d^{in})=\frac{1}{N}\left(q_v+q_ed\right), 
\label{eq:copy2a} 
\end{equation} 
\begin{equation}
G_d=\sum_{d'\geq d}\left(\frac{d'}{d}\right)
\frac{n_{d'}}{N}\left(\frac{q_e}{2}\right)^d\left(1-\frac{q_e}{2}\right)^{d'-d}.
\label{eq:copy2b}
\end{equation}
The first two terms in the right hand side of Eq. (\ref{eq:copy2}) result from the duplication of a
neighbor of a vertex
(with probability $q_ed/N$) and the duplication of a vertex with the creation of an edge between the duplicates (with
probability $q_v/N$), yielding the attachment rate in Eq. (\ref{eq:copy2a}). Moreover, the last three terms are given by
the
divergence of the duplicates, where with probability $n_d/N$ a vertex with degree $d$ is replaced by two duplicates
(factor two in the last two terms).  Thus, the coupling of the duplication and divergence mixes the
equations for different $n_d$. We cannot give an exact derivation of $n_d$ but we can compute the moments of the degree
distribution \cite{vfmv01,kkkr02}. Multiplying Eq. (\ref{eq:copy2}) by $d^l$ and summing over $d$ we
obtain
\begin{equation}
M_l=\sum_dp_dd^l\sim N^{\sigma_l(q_e)},
\label{eq:copy3a}
\end{equation}
where
\begin{equation}
\sigma_l(q_e)=lq_e+2\left[\left(\frac{1+q_l}{2}\right)^l-1\right],
\label{eq:copy3}
\end{equation}
provided $\sigma_l(q_e)>0$. If $\sigma_l(q_e)<0$ the corresponding moment approaches a stationary value
for large $N$. For all $l$ we find a value $q_l$ at which the moments cross from a divergent
behavior to a finite value for $N\to \infty$. In particular for $l=1$ we have $q_1=1/2$ (as obtained
above) and for $l=2$ we obtain $q_2=2\sqrt{3}-3\approx0.46$. Moreover, the nonlinear behavior with $l$ is
indicative of a multi-fractal degree distribution.

\section{Discussion and Conclusions}
\label{sec:conc}

In order to support the analytical calculations, we have performed numerical simulations of the coupled
duplication-divergence
model with graph size ranging from $N=10^3$ to $10^6$. In Fig. \ref{fig3:dd1} we report the generalized exponents
$\sigma_l(q_e)$ as a function of the divergence parameter $q_e$. As predicted by the analytical
calculations, $\sigma_l=0$
at a critical value $q_l$. The general phase diagrams obtained is in good qualitative, but not
quantitative, agreement
with the mean-field predictions and the multi-fractal picture. Noticeably, multi-fractal features are present also in a
recently introduced model of growing networks \cite{dms02} where, in analogy with the duplication process, newly added
vertices inherit the network degree properties from parent vertices.  Multi-fractality, thus, appears related to local
inheritance mechanisms. Multi-fractal distributions have a rich scaling structure where the scale-free behavior is
characterized by a continuum of exponents. This behavior is, however, opposite to usual exponentially bounded
distributions. Even if the evolution rules of the coupled duplication-divergence model are local they introduce an
effective linear preferential
attachment. However, because the edge deletion of duplicate vertices introduce additional heterogeneity in the problem we
obtain a multi-fractal behavior.

The coupling between duplication and divergence is however less relevant to determine the scaling of the average
clustering coefficient with vertex degree. In fact, for the coupled duplication-divergence model Eq. (\ref{eq:copy4}) also
applies, obtaining the inverse proportionality in Eq. (\ref{eq:copy5}). In Fig. \ref{fig3:dd:2} we plot $\left<c\right>_d$
vs. $d$ for different values of $q_e$, manifesting a power law decay but with an exponent larger than 1. With decreasing
$q_e$ (increasing the lost of edges) the power law decay deviates more and more from the predicted
behavior $\left<c\right>_d\sim d^{-1}$. This picture corroborates our hypothesis that if the edge lost is sufficiently
large then a faster decay should be observed. 

On the other hand, the average neighbor degree as a function of the vertex degree for different values of
$q_e$ is despited in Fig. \ref{fig3:dd:3}. Negative degree correlations are manifested by
a power law decay $\left<d_{nn}\right>\sim d^{-0.1}$. The existence of negative degree correlations have
been actually reported in Ref. \cite{ms02} for a protein-protein interaction network. Moreover, a model
based on these correlations have been also proposed in Ref. \cite{blw02}.

\begin{table}
\begin{center}
\begin{tabular}{|l|c|c|}
\hline
Mechanism & $\left<c\right>_d\sim d^{-\beta}$ & $\left<d_{nn}\right>_d\sim d^{\alpha}$\\
\hline
Connecting neighbors & $0<\beta<1$ & $\alpha>0$\\
Random walk & $\beta=1$ & $\alpha\leq0$\\
Duplication-divergence & $\beta\geq1$ & $\alpha<0$\\
\hline
\end{tabular}
\end{center}
\caption{Summary of the correlation properties of the different models analyzed here.}
\label{sum}
\end{table}

After analyzing these models we can conclude that growing networks based on local evolution rules
exhibit an effective linear preferential attachment. The general principle behind it is the following. It
is
true that when we take a vertex at
random the selection does not imply any degree preference, other than the one imposed by the
degree distribution. However, if we take a neighbor of that vertex then some preference is induced. In fact the
probability that vertex $i$ is a neighbor of the randomly selected vertex is simply
\begin{equation}
\frac{d_i}{\sum_j d_j}
\end{equation}
which is exactly the linear preferential attachment considered in the BA model \cite{bajb00a}.
Therefore, the connection to a neighbor of a vertex selected at random leads to an effective linear
preferential attachment.

Another important consequence of the local models considered above is the inverse
proportionality between the average clustering coefficient and the vertex degree, or more
general $\left<c\right>_d\sim d^{-\beta}$. This result is determined by the fact that when a
new edge is created to a vertex then with a certain probability an edge will also be created to
one or more of its neighbors. Thus, locality is again a crucial point. On the other hand, even
if we were not able to find an analytical explanation, these local models are also characterized
by degree correlations among connected vertices.

These features are observed in the three models analyzed here and are summarized in Tab. \ref{sum}. They describe 
different systems such as
technological, social and biological networks, that appear unrelated from the definition of their
evolution rules. The detailed analysis performed here reveals that their main ingredient, they are local
models of growing networks, explains the existence of strong similarities in their topological properties. These
observations can be extended to other local models proposed in the literature. An example is the model
introduced in Ref. \cite{dms01e}, where each time a vertex is added it is connected to both ends of an
edge selected at random. It can be easily shown that this rule also introduces an effective linear
preferential attachment, clustering hierarchy and degree correlations. Another example is the
deactivation model \cite{ke02}, where new vertices are connected to small sub-set of connected
vertices. A detailed study of its topology \cite{vbmpv02} reveals the existence of clustering hierarchy
and degree correlations.

In conclusion, the growing models with local rules exhibit some of the common features of real
graphs. They are characterized by an effective preferential attachment, an
average clustering coefficient that decreases with increasing vertex degree,
and degree correlations. The local knowledge is then a general principle determining the topology of
growing complex networks.

I thank A. Vespignani, Y. Moreno and A.-L. Barab\'asi for helpful comments and discussion.


\end{document}